\documentclass[aps, showpacs, twocolumn]{revtex4-1}

\usepackage{amsmath,amssymb,amsfonts}
\usepackage{tabularx}
\usepackage{epsfig}
\usepackage{graphicx}

\begin{document}

\title{A dynamical dark energy model with a given luminosity distance }

\author{Grigoris Panotopoulos}
\email{Grigoris.Panotopoulos@uv.es}
\date{\today}


\address{Departament de Fisica Teorica, Universitat de Valencia, E-46100 Burjassot, Spain, and
Instituto de Fisica Corpuscular (IFIC), Universitat de 
Valencia-CSIC, Edificio de Institutos de Paterna, Apt. 22085, E-46071, 
Valencia, Spain.}


\begin{abstract}
It is assumed that the current cosmic acceleration is driven by a scalar field, the 
Lagrangian of which is a function of the kinetic term only, and that the luminosity
distance is a given function of the red-shift. Upon comparison with Baryon Acoustic 
Oscillations (BAOs) and Cosmic Microwave Background (CMB) data
the parameters of the models are determined, and then the time evolution of the
scalar field is determined by the dynamics using the cosmological equations. We find
that the solution is very different than the corresponding solution when the non-relativistic 
matter is ignored, and that the universe enters the acceleration era at larger red-shift compared
to the standard $\Lambda CDM$ model.
\end{abstract}


\maketitle

According to our current understanding about the cosmos,
we live in a flat universe which expands in an accelerating
rate and it is dominated by a dark component.
Identifying the origin and nature of dark energy is one of
the biggest challenges for modern cosmology. Since we
can only speculate about what dark energy could be, many
cosmological models have been proposed and studied so
far. The simplest candidate is a cosmological constant with
state parameter $w=-1$, but models with a dynamical
component with an evolving state parameter $w(z)$ also exist
in the literature (for a review on dark energy models see
e.g.~\cite{Copeland:2006wr}). Scalar field models with
a non-canonical kinetic term have been discussed in k-inflation~\cite{ArmendarizPicon:1999rj}, 
in which inflation is not due to the potential, but rather to the kinetic term in the lagrangian, 
and in k-essence models.~\cite{ArmendarizPicon:2000dh, ArmendarizPicon:2000ah}, which are designed 
to address the issue why the cosmic acceleration has recently begun. Very recently, 
in~\cite{Bandyopadhyay:2011dh} 
the authors studied a k-essence model assuming i) a closed form parameterization for the luminosity 
distance (for distances in cosmology see e.g.~\cite{Hogg:1999ad}), proposed previously 
in~\cite{Padmanabhan:2002vv}, and ii) that the
Lagrangian of the scalar field does not depend on the field itself. However, in that work the matter
has been ignored, and in the present article we wish to show that if the presence of matter is also
taken into account, the time evolution of the scalar field is very different. We also find that
according to this model, the universe enters the acceleration era at a larger red-shift compared
to the standard $\Lambda CDM$ model. 

The theoretical framework is basically the same as in~\cite{Bandyopadhyay:2011dh}, apart from the inclusion
of matter in the cosmological equations, and the use of BAOs and CMB data to be mentioned later on. However,
we start by giving a summary of all the ingredients we shall be using, so that the present work is self-contained.
We focus here on the case of a flat universe, since this is a robust prediction of 
inflation~\cite{Guth:1980zm,Lyth:1998xn}, and it is also supported by current data~\cite{Komatsu:2010fb}.
The framework is based on four-dimensional General Relativity coupled to a single scalar field with a general 
Lagrangian $\mathcal{L}(\phi, X)$, where $\phi$ is the scalar field and $X=(\partial \phi)^2/2$ is the 
standard kinetic term. Therefore, our model is described by the action
\begin{equation}
S=\int d^4x \sqrt{-g} \: \left (-\frac{R}{16 \pi G}+\mathcal{L}(\phi, X) \right)
\end{equation}
The energy-momentum tensor for the scalar field is given by
\begin{equation}
T_{\mu \nu}^{(\phi)}=\mathcal{L}_{,X} \partial_{\mu} \phi \partial_{\nu} \phi-\mathcal{L} g_{\mu \nu}
\end{equation}
where $,X$ denotes differentiation with respect to $X$. We can recast this energy-momentum tensor into 
the form of the energy-momentum tensor for a perfect fluid
\begin{equation}
T_{\mu \nu}^{(p.f)}=(\rho + p) u_{\mu} u_{\nu}-p g_{\mu \nu}
\end{equation}
where $\rho, p$ are the energy density and the pressure of the fluid respectively.
The hydrodynamical quantities $\rho, p, u_{\mu}$ are given in terms of $\phi, X$ as follows
\begin{eqnarray}
p & = & \mathcal{L} \\
\rho & = & 2Xp_{,X}-p \\
u_{\mu} & = & \frac{\partial_{\mu} \phi}{\sqrt{2X}}
\end{eqnarray}
Therefore, the equations of motion for the system gravity+scalar field are just the first Friedmann equation 
and the equation for energy conservation
\begin{eqnarray}
H^2 & = & \frac{8 \pi G}{3} \: \rho \\
\dot{\rho} & = & -3H(\rho + p)
\end{eqnarray}
where $H=\dot{a}/a$ is the Hubble parameter, $a$ is the scale factor and the overdot denotes differentiation with 
respect to cosmic time. Defining the sound speed of the scalar field
\begin{equation}
c_s^2 \equiv \frac{p_{,X}}{\rho_{,X}}=\left ( 1+2X \: \frac{p_{,XX}}{p_{,X}} \right )^{-1}
\end{equation}
the equation for energy conservation takes the form
\begin{equation}
\ddot{\phi}+3c_s^2H\dot{\phi}+\frac{\rho_{,\phi}}{\rho_{,X}}=0
\end{equation}
which generalizes the usual Klein-Gordon equation of a canonical scalar field in a 
Friedmann-Robertson-Walker background.

If we now include also the non-relativistic matter, the equations of motion for our system are the 
Friedmann equations and the conservation equation
\begin{eqnarray}
H^2 & = & \frac{\kappa^2}{3} \: \rho \\
\dot{H} & = & -\frac{\kappa^2}{2} \: (\rho+p) \\
0 & = & \dot{\rho}+3H (\rho+p)
\end{eqnarray}
where $\kappa^2=8 \pi G$, $G$ is Newton´s constant, $\rho_m$ is the 
energy density of matter ($p_m=0$), $\rho_X, p_X$ are the energy density and pressure of dark energy respectively, 
$\rho=\rho_m+\rho_X$ is the total energy density and $p=p_m+p_X=p_X$ is the total pressure. Assuming no 
interaction between dark energy and non-relativistic matter, we have two seperate conservation equations
\begin{eqnarray}
0 & = & \dot{\rho_m}+3H \rho_m \\
0 & = & \dot{\rho_X}+3H (\rho_X+p_X)
\end{eqnarray}
where the second equation for dark energy is equivalent to the equation of motion of
the scalar field. If we further assume that the Lagrangian of the scalar field depends
on the kinetic term only, $\mathcal{L}=-F(X)$, the equation of motion can be integrated once 
\begin{equation}
X F_X^2 = C a^{-6}
\end{equation}
where $C$ is an arbitrary integration constant, which later on will be taken to be $\sqrt{C}=1/(8 \pi G)$, 
and $F_X$ is the derivative of $F(X)$ with respect to the kinetic term. All in all, there are three
independent equations, namely the equation of the scalar field, the conservation equation for dust, and
the second Friedmann equation. For a given Hubble parameter as a function of the red-shift $H(z)$, the
time evolution of the scalar field can be found using the set of the basic cosmological equations. This
will be useful later on (see equations (29) and (30), as well as the definitions (27) and (28)).

The comparison of a theoretical model against supernovae
data relies on the minimization of
\begin{equation}
\chi^2=\sum_i \: \frac{(\mu_{th}(z_i)-\mu_{obs}(z_i))^2}{\sigma_i^2}
\end{equation}
with $\mu$ the distance modulus, $\mu=m-M$, where $M$ is the
absolute magnitude and $m$ is the apparent magnitude. The
theoretical apparent magnitude $\mu_{th}$ as a function of the red-shift $z=-1+a_0/a$ is given by
\begin{equation}
\mu(z)=25+5log_{10} \left ( \frac{d_L(z)}{Mpc} \right )
\end{equation}
where the luminosity distance $d_L$ for a flat universe can be
expressed as
\begin{equation}
d_L(z)=c (1+z) \int_0^z \: dx \frac{1}{H(x)}
\end{equation}
and $c$ is the speed of light, while $H(z)$ is the Hubble parameter as a function of 
the redshift.

We also exploit the CMB shift parameter $R$, since it is the least model dependent quantity extracted 
from the CMB power spectrum, i.e. it does not depend on the present value of the Hubble parameter 
$H_0$. The reduced distance $R$ is written as
\begin{equation}
R=(\Omega_m H_0^2)^{1/2}\int_0^{1089} dz/H(z)
\end{equation}
where $\Omega_m$ is today´s value of the matter normalized density, and we use the CMB 
shift parameter value $R = 1.7 \pm 0.03$, as derived in~\cite{Wang:2006ts}.

Independent geometrical probes are Baryon Acoustic Oscillations measurements. Acoustic oscillations in the 
photon-baryon 
plasma are imprinted in the matter distribution. These BAOs have been detected in the spatial 
distribution of galaxies by the SDSS~\cite{Eisenstein:2005su} at a redshift $z = 0.35$. The SDSS team 
reports its BAO measurement in terms of the $A$ parameter
\begin{equation}
A(z=0.35)=D_V(z=0.35) \frac{\sqrt{\Omega_m H^2_0}}{0.35c}
\end{equation}
where the function $D_V(z)$ is defined to be
\begin{eqnarray}
D_V(z) & \equiv & \left(D^2_A(z)\frac{c z}{H(z)}\right)^{1/3} \\
D_A(z) & = & d_L(z)/(1+z) 
\end{eqnarray}
and we have used the value $ASDSS(z = 0.35) = 0.469 \pm 0.017$. We shall assume that the scalar field
Lagrangian, $\mathcal{L}=-F(X)$, is chosen so that the solution of the cosmological equations for the
Hubble parameter $H(z)$ leads to the luminosity distance
\begin{equation}
d_L(z)=\frac{c}{H_0} \: \frac{z (1+\alpha z)}{1+\beta z}
\end{equation}
Then the CMB shift parameter as well as the BAO $A$ parameter, for a given matter density $\Omega_m$, are
certain functions of the two parameters $\alpha, \beta$, and if we require that $R(\alpha, \beta)=1.7$ and
$A(\alpha, \beta)=0.469$ the two parameters can be determined. Taking $\Omega_m=0.29$ we find that the model
is viable as long as the two free parameters take the values
\begin{eqnarray}
\alpha & = & 1.008 \\
\beta & = & 0.318
\end{eqnarray}
and furthermore the distance modulus $\mu$ cannot be distinguished from the corresponding quantity when the matter 
is ignored, see Figure 1 below.

\begin{figure}
\centering
\includegraphics[width=\linewidth]{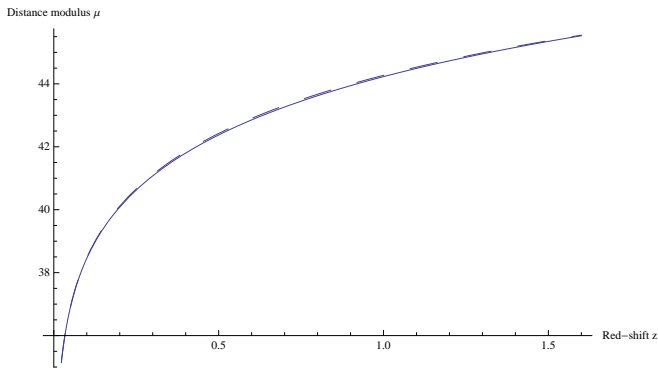}
\caption{Distance modulus versus red-shift for a) a model without matter, $\alpha=1.246$ and $\beta=0.446$ 
(dashed line), and b) our model including matter with $\Omega_m=0.29$, $\alpha=1.008$ and $\beta=0.318$ 
(solid line).}
\end{figure}

To find the time evolution of the scalar field, we introduce the dimensionless scalar field
and time as follows
\begin{eqnarray}
f & = & \frac{\phi}{H_0} \\
\tau & = & H_0 (t-t_0)
\end{eqnarray}
where $t_0$ is the age of the universe today, and we use the second Friedmann equation.
The dimensionless time $\tau$ as a function of the red-shift is given by
\begin{equation}
\tau(z)=-\int_0^z \: \frac{dx}{(1+x)E(x)}
\end{equation}
while the dimensionless field $f$ as a function of the red-shift is given by
\begin{equation} \label{eqn}
f(z)-f(0)=\int_0^z \: dx \left ( \frac{3 \Omega_m}{\sqrt{2} (1+x) E(x)}-\frac{\sqrt{2} E'(x)}{(1+x)^3} \right )
\end{equation}
where $E(z) \equiv H(z)/H_0$, and is given in terms of the assumed (dimensionless) luminosity distance 
$D_L(z)=(H_0/c)d_L(z)$ by
\begin{eqnarray}
E(z) & = & \left ( \frac{d}{dz} \: \left( \frac{D_L(z)}{1+z} \right ) \right )^{-1} \\
D_L(z) & = & \frac{z (1+\alpha z)}{1+\beta z} 
\end{eqnarray}
with $\alpha, \beta$ having the values determined before, and we have used that $\dot{\phi}=\sqrt{2X}$.
Finally, eliminating the red-shift we obtain $f$ as a function of $\tau$, and the solution can be seen in
Figure 2 below. In the same figure we also show the polynomial of second degree that best fits the solution
$f(\tau)$ in the same interval, which is the following
\begin{equation}
P_2(\tau)=0.2385 \tau+0.3297 \tau^2
\end{equation}
We see that the solution exhibits a global minimum at $\tau \simeq -0.35 $, which corresponds to a 
red-shift $z \simeq 0.48$. This is the value for which the two terms in (\ref{eqn}) cancel eachother.

\begin{figure}
\centering
\includegraphics[width=\linewidth]{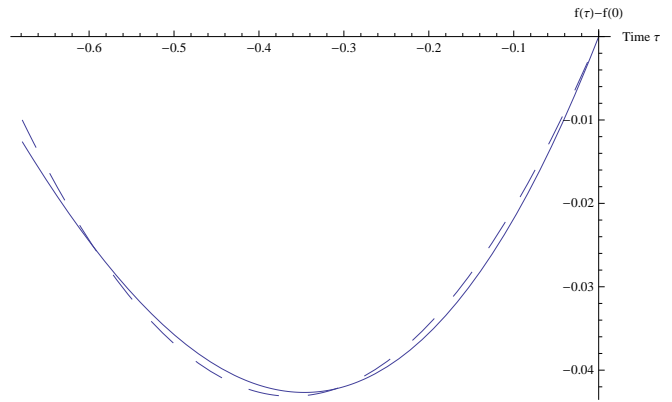}
\caption{Time evolution $f(\tau)$ in our model including matter with $\Omega_m=0.29$, $\alpha=1.008$ and 
$\beta=0.318$ (solid line), and the polynomial of second degree $P_2(\tau)=0.2385 \tau+0.3297 \tau^2$ 
that best fits the solution (dashed line).}
\end{figure}

Before ending our discussion, let us compare the deceleration parameter $q \equiv -\ddot{a}/(aH^2)$ of the
standard $\Lambda CDM$ model with that of our model. Using the definitions, the deceleration parameter
as a function of the red-shift is given by
\begin{equation}
q(z)=-1+(1+z) \frac{H'(z)}{H(z)}
\end{equation}
and can be seen in Figure 3, where for $\Lambda CDM$ we have used the values 
$\Omega_m=0.27, \Omega_{\Lambda}=0.73$. In our model the universe enters into the acceleration era
at $z=1.16$, while in the standard cosmological model this takes place at later times.

\begin{figure}
\centering
\includegraphics[width=\linewidth]{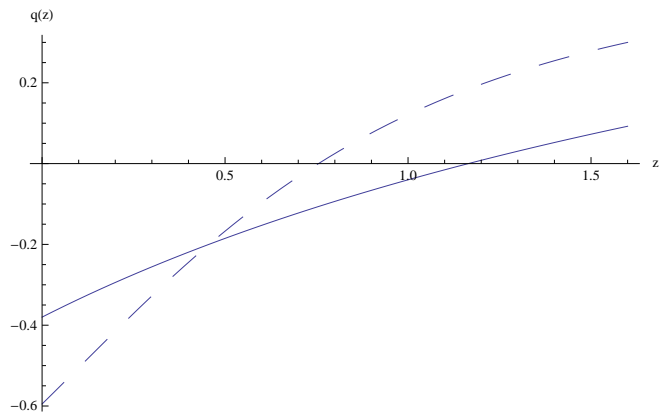}
\caption{Deceleration parameter versus red-shift for a) the standard $\Lambda CDM$ model with 
$\Omega_m=0.27$ and $\Omega_{\Lambda}=0.73$ (dashed line), and b) our model including matter 
with $\Omega_m=0.29$, $\alpha=1.008$ and $\beta=0.318$ (solid line).}
\end{figure}

In summary, in the present work we have assumed that the current cosmic acceleration is driven by a 
scalar field, the Lagrangian of which is a function of the kinetic term only, and that the luminosity
distance is a given function of the red-shift. Upon comparison with Baryon Acoustic 
Oscillations (BAOs) and Cosmic Microwave Background (CMB) data
the parameters of the models are determined, and then the time evolution of the
scalar field is obtained by integrating the cosmological equations numerically. We find
that the solution is very different than the corresponding solution when the non-relativistic 
matter is ignored, and that the universe enters the acceleration era at larger red-shift compared
to the standard $\Lambda CDM$ model.

\section*{Appendix}

Here we briefly discuss how a given function $f(x)$ that vanishes at the origin can be fitted by a 
polynomial of second degree $P_2(x)=ax^2+bx$ in a given interval $x_1 \leq x \leq x_2$. We require
that the deviation of the function from the polynomial is as lower as it can be, or in other words
the integral
\begin{equation}
I(a,b) = \int_{x_1}^{x_2} \: dx (f(x)-P_2(x))^2
\end{equation}
must be extremized. If we set the first derivatives of $I$ with respect to $a, b$ equal to zero
\begin{eqnarray}
\frac{\partial I(a,b)}{\partial a} & = & 0 \\
\frac{\partial I(a,b)}{\partial b} & = & 0
\end{eqnarray}
we obtain a two-by-two system
\begin{eqnarray}
I_3a+I_2b & = & \mu_1 \\
I_4a+I_3b & = & \mu_2
\end{eqnarray}
where we have defined
\begin{eqnarray}
\mu_n & = & \int_{x_1}^{x_2} \: dx \: x^n f(x) \\
I_m & = & \int_{x_1}^{x_2} \: dx \: x^m 
\end{eqnarray}
for $n=1,2$ and $m=2,3,4$. For the given interval and function the above integrals can be computed, and 
solving the simple algebraic system we find the values of the coefficients $a,b$ given in the text.

\section*{Acknowledgments}

The author acknowledges financial support from FPA2008-02878 and Generalitat Valenciana 
under the grant PROMETEO/2008/004.


\end{document}